\documentclass[12pt]{article}
\usepackage{graphicx}
\usepackage{amsfonts}
\usepackage{amsmath,amssymb}
\usepackage{mathrsfs}
\usepackage{setspace}
\usepackage{cite}

\def\beq{\begin{equation}}
\def\eeq{\end{equation}}
\def\bea{\begin{eqnarray}}
\def\eea{\end{eqnarray}}
\def\no{\noindent}
\def\nn{\nonumber}

\def\ba{\begin{array}}
\def\ea{\end{array}}
\def\al{\alpha}

\def\de{\delta}
\def\si{\sigma}
\def\la{\lambda}
\def\La{\Lambda}
\def\th{\Theta}

\def\om{\omega}
\def\ch{\chi}
\def\rh{\rho}
\def\ga{\gamma}

\def\da{\dagger}
\def\l{\langle}
\def\r{\rangle}
\def\v{\vert}
\def\one{1\hskip -1mm{\rm l}}

\begin{document}

\begin{center}
{\Large \bf \sf
     Level density distribution for one-dimensional vertex models \\
      related to Haldane-Shastry like spin chains }

\vspace{1.1cm}

\begin{center}
{\large \em  Dedicated to the memory of Professor Miki Wadati}
\end{center}

\vspace{1.3cm}
{\sf Pratyay Banerjee\footnote{e-mail: pratyay.banerjee@saha.ac.in} and 
B. Basu-Mallick \footnote{Corresponding Author:
e-mail: bireswar.basumallick@saha.ac.in , Phone: +91-33-2337-5346,
FAX: +91-33-2337-4637
}}

\bigskip

{\em Theory Division, Saha Institute of Nuclear Physics, \\
1/AF Bidhan Nagar, Kolkata 700 064, India}

\end{center}

\bigskip

\vspace {1.1 cm}
\baselineskip=16.4pt
\no{\bf Abstract }

The energy level density distributions of some Haldane-Shastry like spin chains 
associated with the $A_{N-1}$ root system have been computed recently by 
Enciso {\it et al.}, exploiting the connection of these spin systems 
with inhomogeneous one-dimensional vertex models whose energy functions 
depend on the vertices through specific polynomials of first or second degree. 
Here we consider a much broader class of one-dimensional vertex models whose
energy functions depend on the vertices through arbitrary polynomials of any 
possible degree. We estimate the order of mean and variance for such energy
functions and show that the level density distribution of all vertex models
belonging to this class asymptotically follow the Gaussian pattern for large
number of vertices. We also present some numerical evidence in support of this
analytical result. 

\vspace{.8 cm}
\no PACS No.: 02.30.Ik, 75.10.Jm, 05.30.-d 

\vspace {.8 cm}
\no Keywords: Haldane-Shastry spin chain, vertex model, Yangian quantum group,  
             level density distribution

\newpage

\no \section{Introduction }
\renewcommand{\theequation}{1.{\arabic{equation}}}
\setcounter{equation}{0}
\medskip

As is well known,  one-dimensional (1D) quantum integrable 
 spin chains can be broadly classified into two 
types depending on their range of interaction. One such class consists of  
spin chains having only local interactions like nearest 
or next-to-nearest neighbour interaction. 
Isotropic and anisotropic versions 
of Heisenberg spin-$\frac{1}{2}$ chain, 
Hubbard model etc. are examples of this type of  
spin models with short range interaction \cite{Su04,KB93,EF05}. 
However, there exist another class of 1D quantum 
integrable spin systems for which  
all constituent spins mutually interact with each other through long range
forces \cite{Su04,Ha96}. 
Haldane-Shastry (HS) spin chain \cite{Ha88,Sh88}, Polychronakos spin chain 
(also known as Polychronakos-Frahm (PF) spin chain) \cite{Po93,Fr93,Po94} 
and Frahm-Inozemstsev (FI) spin chain chain \cite{FI94, BFGR10}
are well known examples of quantum integrable spin systems with long range
interaction. This type of spin chains 
have attracted a lot of attention in recent years due to their exact 
solvability and applicability in a wide range of subjects 
like  generalized exclusion 
statistics and fractional quantum Hall effect \cite{Ha91, Ha93, BGS09, BH09}, 
SUSY Yang-Mills theory and  string theory 
\cite{BDS04,SS04,BS05, BBL09, Se11}, 
Yangian quantum group 
\cite{HHBP92, BGHP93, Hi95, KKN97,  HBM00, Hi00, BBHS07, BBS08} etc.  
For the simplest case of  $A_{N-1}$ root system,  
the Hamiltonians of $su(m)$ invariant, ferromagnetic type 
HS, PF and FI spin chains  
can be written in a general form like
\cite{Ha96,Ha88,Sh88,Po93,Fr93,Po94,FI94, BFGR10}
\beq 
{H} ~=~\sum_{1 \le j<k \le N} ~
\frac{1-P_{jk}}{f(\xi_j-\xi_k)} \, . 
\label{a1} 
\eeq
Here  $\xi_j$ denotes the position of the $j$-th lattice site,
spins can take $m$ possible values on each lattice site,   
${P}_{jk}$ represents the exchange operator which interchanges the spins 
on the $j$-th and $k$-th lattice sites and the strength of this 
exchange interaction is determined by the function
$f(\xi_j-\xi_k)$. It should be noted that, while 
the lattice sites of HS spin chain are uniformly distributed on a circle,
the lattice sites of PF and  FI spin chain
are non-uniformly distributed on a straight line. Supersymmetric 
generalizations of the spin chains (\ref{a1}) have also been 
studied in the literature \cite{Ha93,B99,BMUW99,BB06}.

The difference between quantum integrable models 
with short range interaction and long range
interaction manifests very prominently through 
their connection with classical vertex models. 
It is well known that the 
Hamiltonians of 1D quantum integrable spin chains 
with short range interaction are related to the transfer matrices of some 
exactly solvable 2D vertex models in statistical
mechanics \cite{Su70, Ba71}. However, there exists 
no simple relation between the 
energy levels of such 1D  spin chains and the energy functions of
corresponding 2D vertex models.
On the other hand, a direct connection between
the energy levels of 1D quantum integrable spin chains
with long range interaction and energy functions of 1D vertex models can be 
inferred from some works in the literature \cite{KKN97,Hi00, EFG10}.
Indeed, by applying a rather general technique 
based on Yangian quantum group symmetry, 
it has been shown recently that the energy levels for a class of 
HS like spin chains as well as their supersymmetric
generalizations exactly coincide 
with the energy functions of some
1D inhomogeneous vertex models in classical statistical mechanics \cite{BBK10}. 

The reason behind the close connection between   
1D quantum spin chains with long range interaction
 and 1D classical vertex models
can be traced from the underlying Yangian quantum group symmetry 
of the spin Hamiltonians, which allows one to write down 
their exact spectra in closed forms even
for finite number of lattice sites. 
In the following, we shall briefly recapitulate the consequences of 
Yangian symmetry for the special case of 
nonsupersymmetric $su(m)$ ferromagnetic spin chains. 
A class of irreducible representations of the $Y(sl_{m})$
Yangian algebra, known as `border strips' or  `motifs',   
completely span the Hilbert space of some Yangian 
invariant spin systems with long range interaction
 \cite{Ha93, Ha96, HHBP92, BGHP93, Hi95, KKN97,HBM00,
Hi00, BBHS07, BBS08}.
For the case of a spin system with $N$ number of lattice sites,
boarder strips are denoted by   $\l k_1, k_2, \ldots, k_r \r$,
where $k_i$'s can be taken as any positive integers satisfying the relation 
$\sum_{i=1}^r k_i=N$. Thus the Hilbert space associated with a 
Yangian invariant spin system may be decomposed as
\beq
\mathcal{V}=\sum\limits_{ k_1 + k_2 + \ldots + k_r = N} \oplus \hspace{.2cm} 
{V}_{ \l k_1, k_2, \ldots, k_r \r } \, ,
\label{a2}
\eeq
where ${V}_{ \l k_1, k_2, \ldots, k_r \r }$ denotes
the irreducible vector space associated with 
boarder strip $\l k_1, k_2, \ldots, k_r \r$. Let us restrict 
our attention to a class of Yangian invariant spin systems, 
for which the complete set of energy levels  
can be written in the following way:
\beq
E_{\l k_1, k_2, \ldots, k_r \r}=\sum_{l=1}^{r-1}
\mathcal{E}_N (j)\v_{j=K_l}
\, ,
\label{a3}
\eeq
where $K_l = \sum_{i=1}^l k_i$,  
and $\mathcal{E}_N (j)$ is an arbitrary function 
of the discrete variables $j$ and $N$. 
Due to Yangian symmetry of this class of spin systems,
the multiplicity of the 
eigenvalue $E_{ \l k_1, k_2, \ldots, k_r \r}$ in Eq.~(\ref{a3}) 
coincides with the dimensionality
of the vector space  ${V}_{ \l k_1, k_2, \ldots, k_r \r }$.
Hence, this type of energy eigenvalues are highly degenerate in general. 
 It should be noted that, all energy levels of  
$su(m)$ ferromagnetic HS, PF and FI spin chains 
can be generated through Eq.~(\ref{a3}) as special cases, 
where $\mathcal{E}_N(j)$'s are given by `dispersion relations' of the form 
\cite{HHBP92, BGHP93, Hi95, KKN97,HBM00,Hi00, BBHS07, BBS08, EFG10} 
\beq 
\mathcal{E}_N(j) = \left\{  \begin{array}{ll} j(N-j),  & \mbox{for the HS chain} \\
j, &   \mbox{for the PF chain} \\
j(c+j-1), ~ \mbox{with} ~c >0,  & \mbox{for the FI chain.} 
\end{array} \right. 
\label{a4} 
\eeq

The spectra of all Yangian invariant
spin chains associated with Eq.~(\ref{a3}) 
can be written in an elegant alternative form by using the
energy functions of corresponding
1D inhomogeneous vertex models \cite{BBK10}. Let us consider 
a class of 1D classical vertex models with 
$(N+1)$ number of vertices, which are connected through $N$ number of intermediate bonds. 
Each of these bonds can take one of the $m$ possible states. Therefore,
any state for such vertex models can be represented by 
a path configuration of finite length: 
\beq
\vec{s} \equiv \{ s_1, s_2, \ldots, s_N\},
\label{a5}
\eeq
where $s_i \in \{1,2,\cdots , m \} $ denotes the spin state 
of the $i$-th bond. For constructing vertex models corresponding to  
Yangian invariant spin chains whose energy levels are 
given by Eq.~(\ref{a3}), let us 
define an `energy function' related to the spin path configuration
$\vec{s}$ as 
\beq
E(\vec{s}) = \sum_{j=1}^{N-1} \mathcal{E}_N (j) \de(s_j,s_{j+1}), 
\label{a6}
\eeq
where 
$\de(s_j,s_{j+1})$ is taken as 
\beq
\de(s_j,s_{j+1}) = \left\{ \begin{array}{rl}
0 & \text{if}\hspace{.2cm} s_{j+1} \leq s_{j}
\\
1 & \text{if}\hspace{.2cm} s_{j+1} > s_{j} \, .
\end{array}  \right. 
\label{a7}
\eeq
It should be noted that, the same function $\mathcal{E}_N (j)$ 
has appeared in Eqs.~(\ref{a6})
and (\ref{a3}). 
It can be shown that, 
for all possible functional forms of such $\mathcal{E}_N (j)$,
 the spectrum of the $su(m)$ 
ferromagnetic spin chain generated by eigenvalues (\ref{a3}) 
(along with the degeneracy factors of these energy levels)
completely matches with that of the corresponding vertex model
generated by energy functions (\ref{a6}) \cite{BBK10}.
In particular, the full spectra of 
$su(m)$ ferromagnetic HS, PF and FI spin chains
can be reproduced from the energy function (\ref{a6}), 
by substituting to it the corresponding 
forms of dispersion relations given in Eq.~(\ref{a4}).
Hence, it is possible to extract 
important informations about the  
statistical properties of the spectra of these spin chains
by  analyzing the distribution of  
energy functions for the related 1D vertex models. 
With the help of numerical studies, 
it has been conjectured for a long time that 
the energy level densities for this type of spin chains asymptotically follow 
the Gaussian distribution as the number of lattice sites become very large 
\cite{FG05, EFGR05, BFG09, BFGR08, BFGR08a, BFGR09, BFGR10}.
By using the above mentioned connection with 1D vertex models,  
very recently this conjecture has been proved analytically for
the case of $su(m)$ HS, PF and FI spin chains
associated with the $A_{N-1}$ root system \cite{EFG10}.

In this context, it is interesting to ask whether the Gaussian 
level density distribution of HS, PF and FI spin chains 
is simply a consequence of their 
Yangian quantum group symmetry or this result crucially depends on the
 specific forms of the dispersion relations given in Eq.~(\ref{a4}).
To answer this question, in this article our aim is to explore 
the level density distribution of the 
energy function (\ref{a6}) for a large class of 1D vertex models,  for which  
$\mathcal{E}_N (j)$ can be chosen as an arbitrary polynomial of the
variables $j$ and $N$.  
Even though it may not be possible in general 
to express the Hamiltonians of the spin chains
corresponding to such vertex models 
 in a simple form like (\ref{a1}), 
one can formally express those Hamiltonians in dyadic notation  
through the basis vectors of the Hilbert space $\mathcal{V}$    
in the following way.  
Let us denote the basis vectors of 
 the subspace ${V}_{\l  k_1, k_2, \ldots, k_r \r }$
in Eq.~(\ref{a2}) with the notation
$| \psi_{ k_1, k_2, \ldots, k_r; {\bf{\al}}} \r$, 
where $ {\bf{\al}} \in \{ 1,2, \cdots , \bf{\rh_k} \}$,  and  $\bf{\rh_k}$ 
is the dimension of ${V}_{\l  k_1, k_2, \ldots, k_r \r }$.
The Hamiltonians for the above mentioned class of Yangian invariant 
spin chains may now be written as  
\beq
\mathcal{H} =\sum\limits_{ k_1 + k_2 + \ldots + k_r = N}\hspace{.1cm}
\hspace{.1cm} 
E_{\l k_1, k_2, \ldots, k_r \r}
\sum_{{\bf{\al}}=1}^{\bf{\rh_k}  }| \psi_{ k_1, 
k_2, \ldots, k_r; {\bf{\al}}} \r  \l \psi_{ k_1, k_2, \ldots, k_r; {\bf{\al
}}} | \, ,
\label{a8}
\eeq
where $E_{\l k_1, k_2, \ldots, k_r \r}$ is given by 
Eq.~(\ref{a3}), with $\mathcal{E}_N (j)$ 
being an arbitrary polynomial function of $j$ and $N$.
Hence, in the following,  
our aim is also to explore the level density distribution for   
the class of spin chains represented by  $\mathcal{H}$ in Eq.~(\ref{a8}).  

The arrangement of this article is as follows. In Sec.2 we briefly  
recapitulate the transfer matrix approach of Ref.\cite{EFG10}, 
where it has been found that the level densities for
the case of nonsupersymmetric HS, PF and FI spin chains 
and related 1D vertex models asymptotically follow 
the Gaussian distribution 
as the size of these systems become very large.  
In Sec.3 we show that the above mentioned approach 
can be applied to analytically compute the  
level density distribution for a wide range of 1D vertex models and related 
spin chains, for which the dispersion relations   
are taken as arbitrary polynomials of the variables $j$ and $N$. 
In this section, we also find out the order of mean and variance for 
the energy functions or energy levels  
associated with such dispersion relations. 
In Sec.4 we numerically study the level density distribution
 for some 1D vertex models with different choice of the related 
parameters and find that these numerical results are in complete 
agreement with the analytical prediction of the previous section. 
In Sec.5 we summarize our results  and   
make some concluding remarks
on the possible connection between Yangian quantum group symmetry 
of a spin chain and the asymptotic form of its level density distribution.

\no \section{Level density for known HS like spin chains}
\renewcommand{\theequation}{2.{\arabic{equation}}}
\setcounter{equation}{0}

\vspace{.3cm}
  
In this section, we  review the transfer matrix approach for calculating
the level density distributions in
the cases of ferromagnetic $su(m)$ HS, PF and FI spin chains 
related to the $A_{N-1}$ root system \cite{EFG10}.
A key role in this approach is played by 
the normalised characteristic function of the level density distribution. 
Utilizing the equivalence of these spin chains with 1D vertex models, 
such normalised characteristic function may be defined as 
\beq
\hat{\phi}_N(t) = \l e^{\frac{i t(E(\vec{s}) -\mu)}{\si}} \r   \, , 
\label{b1}
\eeq
where $E(\vec{s})$ is the energy function 
(\ref{a6}) of the  vertex model associated with the spin chain, 
$\mu$ and $\sigma$ denote the mean and the variance of this
energy function respectively, and the notation $\l\mathcal{O} \r$ 
represents the average of $\mathcal{O}$ over all spin path configurations.  
If this characteristic function satisfies the asymptotic relation 
\beq 
\lim_{N \rightarrow \infty}\hat
{\phi}_N(t)=e^{-\frac{t^2}{2}} \,  ,
\label{b2}
\eeq
then it can be shown that \cite{F71} at  $N\rightarrow \infty$ limit 
the normalised level density 
approaches to a Gaussian distribution of the form  
\beq
G(E)=\frac{1}{\sqrt{2\pi\si^2}}e^{-\frac{(E-\mu)^2}{2\si^2}}.
\label{b2a}
\eeq  
Therefore, instead of directly working with the level density 
of a spin chain, one can study the $N\rightarrow \infty$ limit of the related  
characteristic function. Inserting 
the form of $E(\vec{s})$ given in Eq.~(\ref{a6}) to  Eq.~(\ref{b1}), 
it is easy to express $\hat{\phi}_N(t)$ as    
\beq
\hat{\phi}_N(t) 
 = m^{-N} e^{-\frac{i \mu t}{\si} } Z_N ( e^{\frac{ i t}{\si}} ) \, ,
\label{b3}
\eeq
where the `partition function' $Z_N ( e^{\frac{ i t}{\si}} )$ is   
defined through a product of transfer matrices as   
\beq
 Z_N ( e^{\frac{ i t}{\si}} )
=m^{N-1}\sum_{n,n'=1}^m \big[ T(\om_1) T(\om_2) \ldots T(\om_{N-1}) \big]_{nn'} \, .
\label{b4}
\eeq
Here the transfer matrix $T(\om_j)$ is given by 
  $T(\om_j) = T(\om)\v_{\om =\om_j}$, with
\beq
T(\om)=\frac{1}{m}
\begin{pmatrix}
  1& \om^m&\cdots & \om^m\\
  \vdots&\ddots&\ddots&\vdots\\
  \vdots&&\ddots&\om^m\\
  1&\ldots&\ldots&1
\end{pmatrix} \, , 
\label{b5}
\eeq
$\om_j=e^{ i \ga_j t} $   and 
\beq
 \ga_j = \frac{\mathcal{E}_N(j)}{m \si}.
\label{b6}
\eeq
Since this transfer matrix can be diagonalised through an unitary 
transformation,  it may be written as 
\beq
T(\om_j)=U(\om_j) D(\om_j) U^{\da} (\om_j),
\label{b7}
\eeq
where $U(\om_j)$ is an unitary matrix with elements given by 
\beq
  U_{nn'}(\om_j)=
\frac{1}{\sqrt m}\, \big( \om_j e^{\frac{2 \pi i n'}{m}} \big)^{m-n},
\label{b8}
\eeq
and  $ D(\om_j) $ is a diagonal matrix:
\beq
D(\om_j)=\text{diag} \hspace{.1cm} \big(\la_1(\om_j), \ldots, \la_m(\om_j) \big)
\, ,
\label{b9}
\eeq
with
\beq
\la_k(\om_j) =  \frac{1}{m} \sum_{r=0}^{m-1} \left( \om_j
e^{\frac{2 \pi i k}{m}} \right)^r \, . 
\label{b10}
\eeq
With the help of Eqs.~(\ref{b3}), (\ref{b4}) and (\ref{b7}),
one can express $\hat{\phi}_N(t)$ as   
\beq
\hat{\phi}_N(t)
=\frac{e^{-i \mu t/{\si}}}{m} \sum_{n,n'=1}^m M_{nn'}(t),
\label{b11}
\eeq
where the matrix $M$ is given by 
\beq
M(t)=U\left(\om_1\right) 
D\left( \om_1 \right) B_1(t) \cdots D\left( \om_{N-2}\right)
B_{N-2}(t) D\left(\om_{N-1} \right)
U^\da\left(\om_{N-1} \right),
\label{b12}
\eeq
with 
\beq
B_j(t)=U^\da\left( \om_j\right) U\left( \om_{j+1} \right).
\label{b13}
\eeq
It should be observed that,  Eq.~(\ref{b11})
is derived without assuming any specific form  
of the dispersion relation.
Furthermore, for any
functional form of $\mathcal{E}_N (j)$,
the mean and the variance of the energy function $E(\vec{s})$
in Eq.~(\ref{a6}) can be expressed as \cite{EFG10} 
\beq
\mu \equiv \l E(\vec{s}) \r = 
\frac{1}{2}\bigg(1-\frac{1}{m}\bigg)\sum_{j=1}^{N-1} \mathcal{E}_N(j),
\label{b14}
\eeq
and
\beq
\si^2 \equiv  \l \left( E(\vec{s}) -\mu \right)^2 \r =
\bigg(1-\frac{1}{m^2}\bigg)\bigg[\frac{1}{4}
 \sum_{j=1}^{N-1}\mathcal{E}_N(j)^2-
\frac{1}{6}\sum_{j=1}^{N-2}\mathcal{E}_N(j)\mathcal{E}_N(j+1)\bigg].
\label{b15}
\eeq

However, for determining the $N\rightarrow \infty$ limit 
of $\hat{\phi}_N(t)$ in Eq.~(\ref{b11}), it is necessary to choose
some specific form of the dispersion relation.
Inserting the specific forms of dispersion relations
given in Eq.~(\ref{a4})
to Eqs.~(\ref{b14}) and (\ref{b15}), one can explicitly obtain 
$\mu$ and $\sigma^2$ (as some functions of $m$ and $N$) for the 
cases of  HS, PF and FI spin chains. 
By plugging such explicit forms of $\sigma$ in Eq.~(\ref{b6}),
it can be shown that
the asymptotic behaviours of $\ga_j$ and $\ga_j-\ga_{j+1}$ 
for large values of $N$ are given by 
the following relations in the cases of all 
these three spin chains \cite{EFG10}: 
\beq
\ga_j=O(N^{-1/2}) \, ,
\label{b16}
\eeq
and 
\beq
\ga_j-\ga_{j+1}=O(N^{-3/2}) \, ,
\label{b17}
\eeq
 where $j \in \{ 1, 2, \ldots, N-1 \}$ and the notation
$f(N)=O(N^a)$ implies that 
\beq
\lim\limits_{N \rightarrow \infty}| N^{-a}f(N) | = \tau \, ,
\label{b18}
\eeq
with $\tau$ being some non-negative real number. 
With the help of 
Eqs.~(\ref{b8}) and (\ref{b16}),  it is easy to show that
\beq
U \left(\om_j\right)=R+O(N^{-1/2}) \, ,
\label{b19}
\eeq
where $R$ is a constant unitary matrix with elements given by  
\beq
R_{nn'}=\frac{1}{\sqrt{m}} e^{- 2 \pi i n n'/m} \, .
\label{b20}
\eeq
By using 
Eqs.~(\ref{b8}), (\ref{b13}) and (\ref{b17}), one can also determine
the asymptotic form of $ B_j(t) $ as
\beq
B_j(t)=\one+O(N^{-3/2}).
\label{b21}
\eeq
Substituting the relations (\ref{b19}) and (\ref{b21}) to 
Eq.~(\ref{b12}) and taking the $N\rightarrow \infty$ limit, 
one obtains 
\beq
\lim_{N \rightarrow \infty} M(t)=R \La(t )R^{\da},
\label{b22}
\eeq
where $\La(t)$ is a diagonal matrix given by
\beq
\La(t)=\text{diag} \hspace{.1cm} \big(\La_1(t), \ldots, \La_m(t) \big) \, ,
\label{b23}
\eeq
with 
\beq
\La_k(t)=\lim_{N \rightarrow \infty} \prod_{j=1}^{N-1}\la_k(\om_j) \, .
\label{b24}
\eeq
Taking the $N\rightarrow \infty$ limit of Eq.~(\ref{b11}),
and subsequently using Eqs.~(\ref{b22}), (\ref{b20}) and (\ref{b24}),  
it is easy to find that
\beq 
\lim_{N \rightarrow \infty}\hat {\phi}_N(t)=e^{-\frac{it\mu}{\si}} \La_m(t)
=e^{-\frac{it\mu}{\si}} 
\lim_{N \rightarrow \infty} \prod_{j=1}^{N-1}\la_m(\om_j) \, .
\label{b25}
\eeq
By putting $k=m$ in Eq.~(\ref{b10}), one obtains   
$\la_m(\om_j)= \frac{1}{m} \sum_{r=0}^{m-1} e^{irt\gamma_j}$,  
from which it is clear that $\la_m(\om_j) \rightarrow 1$ 
in the limit $ \gamma_j \rightarrow 0 $. 
Expanding $ \ln \la_m(\om_j) $ in a power series of   
 $\gamma_j$, and subsequently
using the expressions (\ref{b6}), (\ref{b14}) and (\ref{b15}) along with the 
asymptotic forms (\ref{b16}) and (\ref{b17}), one gets the relation
\beq
\ln \left\{ \prod_{j=1}^{N-1} \la_m(\om_j)  \right \} 
=\frac{i\mu t}{\si}-\frac{t^2}{2}  \, 
+ O(N^{-1}) \, ,
\label{b26}
\eeq
which in turn leads to 
\beq
\lim_{N \rightarrow \infty} \prod_{j=1}^{N-1}\la_m(\om_j)
=e^{\frac{i \mu t}{\si}-\frac{t^2}{2}}.
\label{b27}
\eeq
Substituting this result to Eq.~(\ref{b25}), one finally obtains the 
desired asymptotic form (\ref{b2}) for the normalised characteristic 
function. Thus it is established that 
the level density distribution for $su(m)$ ferromagnetic 
HS, PF and FI chains asymptotically follow the Gaussian pattern (\ref{b2a})
for large number of lattice sites.

\no \section{Level density for vertex models 
with polynomial type dispersion relations }
\renewcommand{\theequation}{3.{\arabic{equation}}}
\setcounter{equation}{0}

\vspace{.3cm}

From the discussions of the previous section it is evident that 
the asymptotic forms
of $\ga_j$ and $\ga_j-\ga_{j+1}$, given by Eqs.~(\ref{b16})
and (\ref{b17}) respectively, play an important role in determining
the level density distribution for the cases of 
HS, PF and FI spin chains.
In this context it is interesting to observe that,  
even though $\ga_j$ in  Eq.~(\ref{b6}) explicitly depends on 
 the functional form of $\mathcal{E}_N (j)$, 
the asymptotic form of $\ga_j$ (or, $\ga_j-\ga_{j+1}$)
is completely same for all three   
dispersion relations appearing in Eq.~(\ref{a4}). This observation suggests
that some universality might be present in these asymptotic forms
of $\ga_j$ and $\ga_j-\ga_{j+1}$.
As a result, the transfer matrix approach
might be applicable for finding out the level density distribution of 
a much larger class of vertex models and associated quantum
spin chains.

Indeed, in the following our aim is to consider 
a large class of vertex models where 
the dispersion relations can be taken as 
 arbitrary polynomial functions of the variables $j$ and $N$,
and show that the asymptotic forms of $\ga_j$ and $\ga_j-\ga_{j+1}$ are 
universally given by Eqs.~(\ref{b16}) and (\ref{b17}) respectively  
for all such cases. To begin with, we write down the general form of
such polynomial type dispersion relation as 
\beq
\mathcal{E}_N(j)=\sum_{\alpha=0}^r\sum_{\beta=0}^s f_{\alpha, \beta}
\,  N^{\alpha}j^{\beta},
\label{c1}
\eeq  
where  $r$ and $s$ are some 
non-negative integers, and $f_{\alpha, \beta}$'s are arbitrary real numbers.
 For the sake of uniquely expressing a polynomial type  
 dispersion relation in the form (\ref{c1})
with minimum possible values of $r$ and $s$,  
 we assume that (without any loss of generality) there exist 
at least one value of $\beta \in \{ 0, \ldots, s \}$ 
such that $f_{r, \beta} \neq 0$ and at least one value 
of $\alpha \in \{ 0, \ldots, r \} $ such that  $f_{\alpha, s} \neq 0$.
The degree of the two-variable 
polynomial $\mathcal{E}_N(j)$ in Eq.~(\ref{c1}) is denoted by  
$\ch$, which takes an integer value within the range: 
$0 \leqslant \ch \leqslant r+s$. 
Evidently, $f_{\alpha, \beta}$'s appearing in Eq.~(\ref{c1}) 
satisfy the conditions 
\bea
f_{\alpha, \beta} &=& 0 ,~ 
\text{for}\hspace{.2cm} \alpha + \beta > \ch , \nn \\
               &\neq& 0 ,\hspace{.2cm}\text{for at least one pair of}
 ~\alpha, \, \beta~ \text{with}\hspace{.2cm}
\alpha + \beta =
\ch \, .
\label{c2}  
\eea
By using these conditions, it is easy to show that
\beq
\ch \, \geqslant \, max\{r, s\}.
\label{c3} 
\eeq
It may be noted that the general form of dispersion relation (\ref{c1}) yields 
all dispersion relations given in Eq.~(\ref{a4}) as some special cases.
For example, the choice $r=0,~s=1$ along with $f_{0,0}=0,$ $f_{0,1}=1$
gives the dispersion relation for PF spin chain and  
the choice $r=1,~s=2$ along with  
$f_{0,0}=f_{0,1}= f_{1,0}=f_{1,2}=0,  $ 
$f_{0,2}=- f_{1,1}= -1$
gives the dispersion relation for HS spin chain. Clearly, 
$\chi=1$ for the vertex model corresponding to PF spin chain and 
$\chi=2$ for the vertex models corresponding to HS and FI spin chains. 

Even though it is easy to find out the explicit form of $\sigma^2$ 
by using Eq.~(\ref{b15}) for the above mentioned type of special cases
 associated with small values of $r$ and $s$,
it is quite cumbersome to do so for the dispersion relation (\ref{c1})
with high values of $r$ and $s$. 
Therefore, instead of trying to compute the exact form of $\sigma^2$ explicitly, 
at present our strategy is to estimate the order of 
$\sigma^2$ for the general form of dispersion relation (\ref{c1}).
To this end,
we consider the `big $\Theta$ notation' for the order of a function:
$f(N)=\th(N^{a})$, which implies that
\beq 
\lim\limits_{N \rightarrow \infty} | N^{-a}f(N)|=\tau,
\label{c4}
\eeq
where $\tau$ is a positive real number. It should 
be observed that, in contrast to the case of
the `big O' notation defined through   
 Eq.~(\ref{b18}), at present $\tau$ is not allowed to take the zero value.
As a result,  the expression $f(N)=\th(N^a)$ carries more 
precise information about the asymptotic form of $f(N)$
in comparison with the expression $f(N)=O(N^a)$.
Indeed, while $f(N)=O(N^a)$ implies that $f(N)$ is bounded above by 
$N^a$ (up to a constant factor) asymptotically, 
$f(N)=\th(N^a)$ implies that $f(N)$ is bounded above and below by $N^a$
(up to constant factors)  asymptotically \cite{B58, JJ00,K76}.
In the following, we shall use both of these notations for 
the order of a function according to our convenience.  

At first, let us examine the order of $\mathcal{E}_N(j)$ in two
possible cases: i) $j=\th(1)$, where $N\rightarrow \infty$ limit
is taken by keeping a fixed value of $j$,  
 and ii)  $j=\th(N)$, where $N\rightarrow \infty$ limit
is taken in such a way that  $j/N$ tends to a fixed value.
Since there exists 
at least one value of $\beta \in \{ 0, \ldots, s \}$ 
for which $f_{r, \beta} \neq 0$ in (\ref{c1}), we easily obtain
the order of $\mathcal{E}_N(j)$
for case i) as  
\beq
\mathcal{E}_N(j)=\th(N^{r}) \, . 
\label{c5}
\eeq
Next we consider the case ii), which is defined through the 
limiting procedure  
\beq
\lim_{N\rightarrow \infty}\, \frac{j}{N} \, = \, \rh \, ,
\label{c6} 
\eeq  
where $\rho$ is a real parameter taking value 
within the range: $0 < \rh\leqslant 1$. Due to Eq.~(\ref{c6}), 
one may write $\mathcal{E}_N(j) \approx \mathcal{E}_N(\rho N)$
for large values of $N$.
Substituting $\rho N$ in the place of $j$ in Eq.~(\ref{c1}), we obtain
\beq
\mathcal{E}_N(\rh N)
=\sum_{\alpha=0}^r\sum_{\beta=0}^s 
f_{\alpha, \beta}\hspace{.1cm} \rh^{\beta} N^{\alpha + \beta}.
\label{c7} 
\eeq  
Let us define a new variable as $\xi \equiv \alpha + \beta$. 
Making a change of the summation variables in 
Eq.~(\ref{c7}),  it can be expressed as 
\beq
\mathcal{E}_N(\rh N)
=\sum_{\xi=0}^{r+s} \, 
\sum_{\beta=max\{ 0, \xi-r \}}^{min\{\xi, s\}} \, 
f_{\xi - \beta , \beta}\,  \rh^{\beta} N^{\xi}.
\label{c8} 
\eeq  
Collecting the coefficients
of $N^{\xi}$ in the above equation, we get 
\beq
\mathcal{E}_N(\rh N)=\sum_{\xi=0}^{r+s} g_{\xi}(\rh) N^{\xi},
\label{c9} 
\eeq   
where $g_{\xi}(\rh)$ is a polynomial of $\rho$ given by 
\beq
g_{\xi}(\rh)=\sum_{\beta=max\{ 0, \xi-r \}}^{min\{\xi, s\}} f_{\xi-\beta,
\beta}\hspace{.1cm} \rh^{\beta}.
\label{c10} 
\eeq
Since from Eq.~(\ref{c2}) it follows that $g_{\xi}(\rh)=0$ for $\xi > \ch$,
one can rewrite Eq.~(\ref{c9}) as 
\beq
\mathcal{E}_N(\rh N)=\sum_{\xi=0}^{\ch} g_{\xi}(\rh) N^{\xi}.
\label{c11} 
\eeq    
With the help of Eqs.~(\ref{c6}) and (\ref{c11}), we obtain 
\beq
 \lim_{N\rightarrow \infty} \, \v N^{-\chi} \, 
 \mathcal{E}_N(j) \v \, =
\lim_{N\rightarrow \infty} \, \v N^{-\chi} \, 
 \mathcal{E}_N(\rh N) \v \, = \, \v g_{\chi}(\rho) \v \, .
\label{c12} 
\eeq  
Since $g_{\ch}(\rh)$ is a polynomial of $\rho$, it is bounded 
within the range $0 < \rh \leqslant 1 $.
If $g_{\ch}(\rh) \neq 0$ for some value of $\rho$, then by 
using Eqs.~(\ref{c4}) and (\ref{c12}) we obtain the order of 
 $\mathcal{E}_N(j)$ as
\beq
\mathcal{E}_N(j) = \th(N^{\ch}). 
\label{c13} 
\eeq 
On the other hand, 
if $g_{\ch}(\rh) = 0$ for some value of $\rho$, 
then  Eq.~(\ref{c11}) reduces to 
$\mathcal{E}_N(\rh N)=\sum_{\xi=0}^{\tilde{\ch}} g_{\xi}(\rh) N^{\xi}$, 
where $\tilde{\ch}<\ch$ and $g_{\tilde{\ch}}(\rh) \neq 0$. 
Hence, for the case $g_{\ch}(\rh) = 0$, we obtain
the order of $\mathcal{E}_N(j)$ as 
\beq
\mathcal{E}_N(j) = \th(N^{\tilde{\ch}}) \, , 
\label{c14} 
\eeq  
where $\tilde{\ch}<\ch$. 
Replacing $\xi$ by $\ch$ in Eq.~(\ref{c10}) and using Eq.~(\ref{c3}), 
one can get an explicit expression for $g_{\ch}(\rh)$ as 
\beq
g_{\ch}(\rh)=\sum_{\beta= \ch-r}^s 
f_{\ch-\beta, \beta}\hspace{.1cm} \rh^{\beta}. 
\label{c15} 
\eeq
Since $g_{\ch}(\rh)$ is a polynomial in $\rh$ of degree less than or
equal to $s$, it can have at
most $s$ number of distinct zeros within the interval $0 < \rh \leqslant 1 $.
For such zero points of $g_{\ch}(\rh)$, the order of  $\mathcal{E}_N(j)$
is evidently given by Eq.~(\ref{c14}). 
However, the zero points of $g_{\ch}(\rh)$ clearly form a set of zero measure
within the range $0 < \rh \leqslant 1 $,
and, therefore,  the order of  $\mathcal{E}_N(j)$
is given by Eq.~(\ref{c13}) for generic values of $\rho$.

Let us now compare the orders of $\mathcal{E}_N(j)$  given in 
 Eqs. (\ref{c5}), (\ref{c13}) and (\ref{c14}) for various possible cases. 
Since due to Eq.~(\ref{c3}) it follows that $\chi\geq r$,
the order of $\mathcal{E}_N(j)$ given by (\ref{c13}) is clearly 
the highest one among all of these cases. 
Therefore, at $N\rightarrow \infty $ limit, 
the expression of $\sigma^2$ in Eq.~(\ref{b15}) 
would get dominant contribution 
from those values of $j$ which satisfy the 
conditions  $j=\th(N), ~g_{\ch}(\rh) \neq 0$ and  
lead to Eq.~(\ref{c13}).
 
Next we explore the stability condition for the order of 
$\mathcal{E}_N(j)$ given in Eq.~(\ref{c13}),
under the variation of parameter $j$ (or, equivalently, $\rho$). 
Since the polynomial
$g_{\ch}(\rh)$ can have at most $s$ number of distinct zeros
within the interval $0 <\rh \leqslant 1 $,
we can always find two distinct numbers $\rh_1$ and $\rh_2$ 
within this interval 
such that $g_{\ch}(\rh) \neq 0$ for 
$\rh_1 \leqslant \rh \leqslant \rh_2$ and 
\beq
\rh_2-\rh_1 = \frac{1}{s+1} \, .
\label{c16} 
\eeq
Due to Eq.~(\ref{c6}), one can write $N \rh_1 \approx j_1$ and
$N \rh_2 \approx j_2$ for large values of $N$,
 where $j_1$, $j_2$ are some integers. 
Since $g_{\ch}(\rh) \neq 0$ for $\rh_1 \leqslant \rh \leqslant \rh_2$, 
it is clear that Eq.~(\ref{c13}) is satisfied for any $j \in \{j_1,
j_1+1, \ldots, j_2 \}$.  
Thus, we can write
\beq
\mathcal{E}_N(j_1 + \vartriangle \! j)= \th(N^{\ch})  \, ,
\label{c17} 
\eeq
where ${\vartriangle \! j}$ may  
take any value within the range $ 0 \leqslant \vartriangle \! j 
\leqslant j_2-j_1 $.
By using (\ref{c16}), 
we obtain $ j_2-j_1 \approx  N(\rho_2-\rho_1) =
 \frac{N}{s+1}$, which yields the order of $ j_2-j_1$ as 
\beq
j_2-j_1 = \th(N) \, .
\label{c18}
\eeq
Eqs.~(\ref{c17}) and (\ref{c18}) imply that, there exists
at least one value of $j$ (denoted by $j_1$ here)
for which $\mathcal{E}_N(j) = \Theta (N^\chi)$
and this asymptotic behaviour of $\mathcal{E}_N(j)$
 is preserved for a variation of 
$j$ of order $N$. This type of stability in the asymptotic
behaviour of $\mathcal{E}_N(j)$ will play an important role 
in our analysis in finding out the order of $\sigma^2$.  

For the purpose of finding out the asymptotic behaviour of $\sigma^2$,
it is also needed to calculate the order of \hspace{.1cm}
$\mathcal{E}_N(j+1) - \mathcal{E}_N(j)$. 
By using Eq.~(\ref{c1}), we obtain
\begin{equation*}
\mathcal{E \hskip .02 cm  '}_N(j) \equiv
\mathcal{E}_N(j+1) - \mathcal{E}_N(j)=\sum_{\alpha=0}^r
\sum_{\beta=1}^s\sum_{\nu=0}^{\beta-1} \, 
{^\beta} {C_\nu}f_{\alpha, \beta}\hspace{.1cm}  N^{\alpha} j^{\nu} \, ,
\end{equation*}
where ${^\beta} {C_\nu} \equiv \frac{\beta !}{\nu ! (\beta - \nu)!}$. 
Interchanging the  order of summations over $\beta$ and $\nu$
in the above equation, and renaming these summation variables,
we express $\mathcal{E \hskip .02 cm  '}_N(j) $  in exactly same form 
as $\mathcal{E}_N(j)$ 
in Eq.~(\ref{c1}): 
\beq
\mathcal{E \hskip .02 cm  '}_N(j) 
=\sum_{\alpha=0}^{r'} \sum_{\beta=0}^{s'}
{f'}_{\alpha, \beta}
\hspace{.1cm} N^{\alpha} j^{\beta},   
\label{c19} 
\eeq
where $r' \leq r$, $s'=s-1$ and  ${f'}_{\alpha, \beta}$ is given by
\beq
 {f'}_{\alpha, \beta}=
\sum_{\nu = \beta + 1}^s \, f_{\alpha, \nu} \, 
 {^{\nu}} C_{\beta} 
 \, .
\label{c20} 
\eeq
Consequently, our analysis   
for finding out the order of 
$\mathcal{E}_N(j)$ would also be applicable for the case of 
$\mathcal{E \hskip .02 cm  '}_N(j)$. Let us denote the degree 
of $\mathcal{E \hskip .02 cm  '}_N(j)$ in Eq.~(\ref{c19}) 
by $\chi'$. With the help of Eq.~(\ref{c20}) and 
the analog of Eq.~(\ref{c2}) for the present case, 
it is easy to find that
\beq
{\ch'} \leqslant \ch - 1 .  
\label{c21} 
\eeq
In analogy with the polynomial $g_{\chi}(\rho)$ in Eq.~(\ref{c15}), 
one can define the polynomial ${\tilde g}_{\chi'}(\rho)$
for the present case as 
\beq
{\tilde g}_{\ch'}(\rh)=\sum_{\beta= \ch'-r'}^{s'} 
f'_{\ch' -\beta, \beta}\hspace{.1cm} \rh^{\beta}. 
\label{c22} 
\eeq
Proceeding in the same way as we have done to
find out the maximum possible order of $\mathcal{E}_N(j)$ in 
Eq.~(\ref{c13}), 
it is easy to show that the order of $\mathcal{E \hskip .02 cm  '}_N(j)$
becomes maximum when $j$  satisfies the relations 
$j=\th(N)$ and ${\tilde g}_{\ch'}(\rh) \neq 0$, and 
this order 
of $\mathcal{E \hskip .02 cm  '}_N(j)$ 
is given by 
\beq
\mathcal{E \hskip .02 cm  '}_N(j)= \th(N^{\ch'}) \, .
\label{c23} 
\eeq  


Next, to calculate the order of $\si^2$ 
for any dispersion relation of the form (\ref{c1}), 
we rewrite Eq.~(\ref{b15}) as 
\beq
\sigma^2=I_1 + I_2 + I_3,
\label{c24}
\eeq
where
\begin{subequations}
\label{c25}
\beq
\label{c251}
I_1 =
 \frac{1}{12} \bigg(1-\frac{1}{m^2}\bigg) \sum_{j=1}^{N-1}
\mathcal{E}_N(j)^2, 
\eeq
\beq
\label{c252}
\hspace{1cm} I_2 = - \frac{1}{6} \bigg(1-\frac{1}{m^2}\bigg) \sum_{j=1}^{N-2}
\mathcal{E}_N(j) \mathcal{E \hskip .02 cm  '}_N(j)  ,
\eeq
\beq
\label{c253}
I_3 = \frac{1}{6} \bigg(1-\frac{1}{m^2}\bigg) \mathcal{E}_N(N-1)^2. 
\eeq
\end{subequations}
Since terms like  
$\mathcal{E}_N(j)^2\, $ appearing in the r.h.s of (\ref{c251})
are positive for all values of $j$, their contribution in $I_1$
 can not cancel out each other. Therefore, 
 by applying Eqs. (\ref{c17}) and (\ref{c18}), we find that  
that $I_1 = \th(N^{2\ch + 1})$. 
Due to Eqs.~(\ref{c21}) 
and (\ref{c23}), it is evident that the 
order of $I_2$ in (\ref{c252}) must be less than that of $I_1$. 
Moreover, since the expression of $I_3$ in (\ref{c253})
does not contain any summation variable, 
the order of it must be less than that of $I_1$. 
Consequently, the dominant contribution  
in the r.h.s. of Eq.~(\ref{c24}) comes from the term $I_1$. 
Thus, we obtain the order of $\si^2$ as
\beq
\si^2=\th(N^{2\ch+1}).
\label{c26}
\eeq
Since the parameter $\tau$ in Eq.~(\ref{c4}) takes a nonzero value, 
it is possible to `invert' the relation (\ref{c26}) and 
calculate the order of $1/\sigma$ as   
\beq
\frac{1}{\si}=\th(N^{-\ch - \frac{1}{2}}).
\label{c27}
\eeq

Now, we are in a position to find out the orders of  
$\ga_j$ and $\ga_j-\ga_{j+1}$
for any polynomial type dispersion relation of the form (\ref{c1}).
%
Let us first consider the case for which $j$ satisfies the relations:  
$j=\th(N)$, $g_{\ch}(\rh) \neq 0$, and 
the order of $\mathcal{E}_N(j)$
is given by Eq.~(\ref{c13}). 
By using Eqs.~(\ref{b6}), (\ref{c13}) and (\ref{c27}), we obtain the 
order of $\ga_j$ for this case as
\beq
\ga_j=\th(N^{-\frac{1}{2}}) \, .
\label{c28}
\eeq
Since the order of $\mathcal{E}_N(j)$ becomes maximum 
for the above mentioned case,
it is evident that
the order of $\ga_j$ can not exceed the value 
given in the r.h.s of Eq.~(\ref{c28})
for any possible choice of $j$. In other words, 
the asymptotic form of $\ga_j$ at $N\rightarrow \infty$ limit 
can not exceed $N^{ -\frac{1}{2}}$ (up to some multiplicative factor) 
for any possible choice of $j$.   
Hence,
by using the `big O' ordering defined in Eq.~(\ref{b18}),  
we obtain a relation valid for any value of $j$ as 
\beq
\ga_j=O(N^{-\frac{1}{2}}) \, . 
\label{c29}
\eeq
Next, for calculating the order of 
$\ga_j-\ga_{j+1}$, we proceed in exactly same way and consider  
the case for which $j$ satisfies the relations: 
$j=\th(N)$, ${\tilde g}_{\ch'}(\rh) \neq 0$, and leads to the 
maximum order of 
$\mathcal{E \hskip .01 cm  '}_N(j)$ given by Eq.~(\ref{c23}).  
Using Eqs.
(\ref{b6}), (\ref{c23}) and (\ref{c27}),  we obtain the 
order of $\ga_j-\ga_{j+1}$ for this case as
\beq
\ga_j-\ga_{j+1}=
\th(N^{\chi'-\chi -\frac{1}{2}}) \, .
\label{c30}
\eeq
In analogy with Eq.~(\ref{c28}), the above equation implies that
the asymptotic form of $\ga_j-\ga_{j+1}$
can not exceed $N^{\chi'-\chi -\frac{1}{2}}$  
 for any possible choice of $j$. 
Moreover, due to Eq.~(\ref{c21}), it follows that 
$\chi'-\chi -\frac{1}{2} \leq - \frac{3}{2}$. As a result, 
we can express the order of  $\ga_j-\ga_{j+1}$ for any
value of $j$ as  
\beq
\ga_j-\ga_{j+1}=
O(N^{-\frac{3}{2}}) \, .
\label{c31}
\eeq

Thus we are able to prove that, 
for any polynomial type dispersion relation of the form (\ref{c1}),
the asymptotic behaviours of $\ga_j$ and $\ga_j-\ga_{j+1}$ 
are given by Eqs.~(\ref{c29}) and (\ref{c31}) respectively. 
Since these asymptotic forms do not depend at all 
on the parameters or coefficients present in the 
dispersion relation (\ref{c1}), they are indeed  
very universal in nature. 
Consequently, by applying the transfer matrix approach exactly like the
cases of HS, PF and FI spin chains as have been discussed in Sec.2, 
we can show that the level density for 
all 1D vertex models and quantum spin chains
associated with the polynomial type dispersion relation (\ref{c1}) 
asymptotically follow the Gaussian distribution (\ref{b2a})
at $N\rightarrow \infty $ limit. 

It is worth noting that, as a by product of our analysis
on the level density distribution, we obtain  
 Eq.~(\ref{c26}) which gives a general expression 
for the order of $\sigma^2$
 for any dispersion relation of the form (\ref{c1}).   
In this context,  it is interesting to ask whether such a formula exists 
for the asymptotic form of mean energy $\mu$.
Since the sign of the term $\mathcal{E}_N(j)$ in Eq.~(\ref{b14}) 
may change for different choice of $j$, 
such terms may cancel out each other in the leading order.
Due to this reason, it is difficult to get a general formula
for the  asymptotic form of $\mu$ through the `big $\Theta$' ordering  
of a function. Nevertheless,
by using Eqs.~(\ref{c17}) and (\ref{c18}), we derive a completely general
formula for the upper limit of the  asymptotic form of $\mu$ as
\beq
\mu=O(N^{\ch + 1}).
\label{c32}
\eeq 
Let us now restrict our attention to those  
dispersion relations of the form (\ref{c1}), for which  
$\mathcal{E}_N(j)$'s in Eq.~(\ref{b14}) do not completely cancel
out each other in the leading order. For such cases, 
the asymptotic form of $\mu$ can be expressed through the 
more precise counterpart of Eq.~(\ref{c32}) given by 
\beq
\mu=\th(N^{\ch + 1}).
\label{c33}
\eeq
In particular,  
if the polynomial $g_{\ch}(\rh)$
in Eq.~(\ref{c15}) does not vanish within the range $0 < \rh < 1$, then
all $\mathcal{E}_N(j)$'s have the same sign  
in the leading order and they can not cancel out each other. 
Therefore, the 
asymptotic form of $\mu$ would be given by Eq.~(\ref{c33})
for this case. By using Eqs.~(\ref{a4}) and (\ref{c15}) 
we find that, for the cases of PF, HS and FI spin chains
 $g_\chi(\rho)$ is given by $\rho$, 
$\rho(1-\rho)$ and $\rho^2$ 
respectively. Since these functions of $\rho$ do not vanish  
within the range $0 < \rh < 1$, Eq.~(\ref{c33}) is applicable
for all of the corresponding spin chains. We have already noted that, 
$\chi=1$ for the case of PF spin chain and 
$\chi=2$ for the cases of HS and FI spin chains. 
Substituting these values of $\chi$ to Eqs.~(\ref{c26}) and (\ref{c33}), 
one can very easily reproduce the known results like 
$\sigma^2=\th(N^{3})$, $\mu=\th(N^{2})$ for PF spin chain and 
$\sigma^2=\th(N^{5})$, $\mu=\th(N^{3})$ for HS and FI spin chains. 
 
\no \section{Numerical results for a vertex model}
\renewcommand{\theequation}{4.{\arabic{equation}}}
\setcounter{equation}{0}
\vspace{.3cm}

In the previous section we have analytically shown that 
the level density for a large class of 1D vertex models 
and Yangian invariant quantum spin chains, 
associated with polynomial type dispersion relations of the
form (\ref{c1}), 
asymptotically follow the Gaussian distribution (\ref{b2a}).
Here our aim is to compare this result with numerical calculations, 
by choosing some specific form of the dispersion relation (\ref{c1})
and taking different possible values for the related parameters.

The numerical data for the level density distribution
corresponding to any given dispersion relation can be 
obtained from the energy function $E(\vec{s})$ 
defined in Eq.~(\ref{a6}). In general, one can 
choose the spin path configuration $\vec {s}$
in many possible ways so that they lead to the same value of $E(\vec{s})$, 
say, $E_i$. Let us denote the number 
of such different spin path configurations by $D^{(m)}(E_i)$.  
In the case of Yangian invariant quantum spin chains 
associated with the vertex models, 
this $D^{(m)}(E_i)$ represents the degeneracy factor of
the energy level $E_i$.  Choosing the values of $N$ and $m$ 
within some range, and using a symbolic software package like Mathematica, 
it is  possible to obtain the values of all $E_i$ and $D^{(m)}(E_i)$
for a given dispersion relation. With the help of such numerical data,  
one can draw the histogram for the corresponding level density and compare
it with the Gaussian distribution. Furthermore, for the purpose of eliminating
the effect of local fluctuations in the level density distribution, 
one may use the above mentioned numerical data to calculate the  
cumulative level density given by
\beq
F(E)=\frac{1}{m^N} \sum_{E_i\leq E} \mathcal{D}^{(m)} (E_i) \, ,
\label{d1}
\eeq
and check whether this $F(E)$ agrees well 
with the error function:  
\beq
C(E)=\frac{1}{\sqrt{2\pi\si^2}} \int_{-\infty}^E e^{-\frac{(x-\mu)^2}{2\si^2}} dx =
\frac{1}{2}\left[1+\rm{erf}\left(\frac{E-\mu}
{\surd{2}\sigma}\right)\right]. 
\label{d2}
\eeq

To get some specific dispersion relation with rather simple form, 
let us choose $r=1$ and
$s=2$ in Eq.~(\ref{c1}), and set the values of three coefficients viz. 
$f_{0, 0}$, $f_{1, 0}$ and $f_{1,2}$ equal 
to zero. Thus, we obtain 
\beq
\mathcal{E}_N(j)=f_{0, 1} j + f_{0, 2} j^2 + f_{1, 1} N j \, , 
\label{d3}
\eeq 
where $f_{0, 1} $ is an arbitrary real parameter, while  
$f_{0, 2}$ and $f_{1, 1}$ are  nonzero  
real parameters. Note that the degree of the polynomial $\mathcal{E}_N(j)$  
in Eq.~(\ref{d3}) is given by $\chi=2$.
It may be observed that, by putting $f_{0, 1}=0$, $f_{0,
2}=-f_{1,1}=-1$ in Eq.~(\ref{d3}), one obtains the  
dispersion relation for the HS spin chain. 
Thus the dispersion relation for the HS spin chain emerges as
a special case of the dispersion relation (\ref{d3}).
%

 
For the purpose of comparing level density with the 
Gaussian distribution (\ref{b2a}), it is necessary to find out 
the values of corresponding mean energy and variance.
Substituting $\mathcal{E}_N(j)$ in (\ref{d3})  
to Eqs.~(\ref{b14}) and (\ref{b15}) respectively, we obtain 
 explicit expressions for the mean energy and variance as 
\begin{subequations}
\label{d4}
\beq
\mu=\frac{1}{12}\bigg(1-\frac{1}{m}\bigg)
 N (N-1)\hspace{.1cm}[\hspace{.05cm}(2f_{0,2}+3f_{1, 1})
\hspace{.05cm}N+(3f_{0, 1}-f_{0, 2})\hspace{.05cm}] \, ,
\label{d4a}
\eeq
\beq
\sigma^2=\frac{1}{360}\bigg(1-\frac{1}{m^2}\bigg)
N(N-1)\hspace{.1cm}(AN^3+BN^2+CN+D) \, , \hspace{.83cm}
\label{d4b}
\eeq
\end{subequations}
where
\begin{eqnarray*}
&&\hspace{-.4cm}A=6f_{0, 2}^2+5f_{1, 1}(2f_{1,1} +3f_{0, 2}), ~
B=3f_{0, 2}(7f_{0, 2}+5f_{0,1}) +5f_{1, 1}(5f_{1, 1} +
9f_{0, 2}+ 4f_{0,1}), \nn \\ 
&&\hspace{-.4cm}C=5f_{0, 1}(2 f_{0, 1} \!  + 9f_{0, 2} 
 \! +10f_{1, 1} ) \! - 49f_{0, 2}^2 \! -60f_{0, 2}f_{1, 1}, \,   
D=5f_{0, 1} (5f_{0, 1} \!  -12f_{0, 2}) \!  + 11f_{0, 2}^2. \nn 
\end{eqnarray*}
Using Eq.~(\ref{d4b}), we find that
\begin{equation}
\lim_{N \rightarrow \infty}
\frac{\sigma^2}{N^5}=\frac{1}{360}\bigg(1-\frac{1}{m^2}\bigg)(6f_{0, 2}
^2+10f_{1, 1}^2+15f_{0, 2}f_{1, 1}).
\label{d5}
\end{equation}
It is interesting to note that the factor $(6f_{0, 2}
^2+10f_{1, 1}^2+15f_{0, 2}f_{1, 1})$ appearing in the r.h.s. 
of the Eq.~(\ref{d5}) can not vanish for any real  
nonzero  values of the parameters 
$f_{0, 2}$ and $f_{1, 1}$.  Thus from Eq.~(\ref{d5}) it follows that  
$\si^2=\Theta(N^5)$ for all allowed values of the related 
parameters, and this result is consistent with 
the general relation (\ref{c26}) for $\chi=2$. 
On the other hand, by using (\ref{d4a}), it is easy to see  that 
the asymptotic form of $\mu$ depends heavily on the specific choice 
of the coefficients. For example,  
$\mu=\Theta(N^3)$ for the case $2f_{0,2}+3f_{1, 1}\neq 0$, 
and the order of $\mu$ would be lower than $N^3$ otherwise.
Again these observations are consistent with the general relation
(\ref{c32}) for $\chi=2$.
 
As a specific case, let us choose $m=2$ and $N=40$,
and investigate the nature of level density distribution for 
the dispersion relation (\ref{d3}) by taking the related parameters 
as $f_{0, 1}=1$, $f_{0, 2}=-1$, $f_{1, 1}=1$.  By using Mathematica,
we get the numerical values of all $E_i$ and $D^{(m)}(E_i)$ for this 
case and plot a suitable histogram in Fig.$1$.
\begin{figure}[htb]
\begin{center}
\resizebox{100mm}{!}{\includegraphics{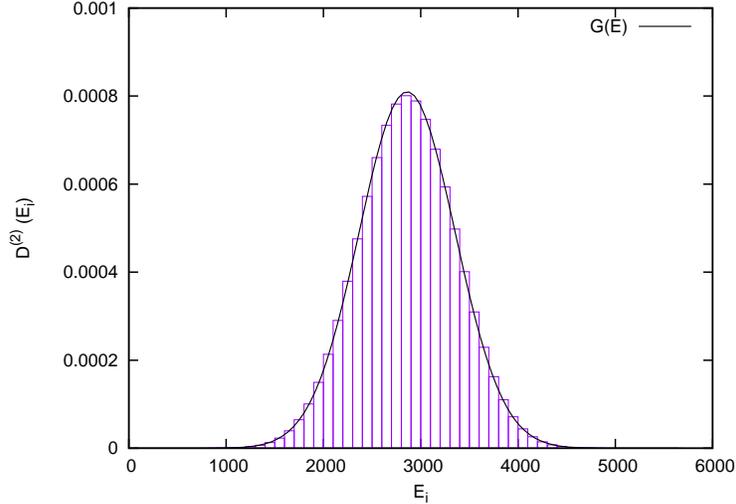}}
{\small{\caption{Continuous curve represents the Gaussian distribution 
$G(E)$ and the histogram 
represents the normalized level density distribution for 
the dispersion relation (\ref{d3}) with coefficients 
$f_{0, 1}=1$, $f_{0, 2}=-1$, $f_{1, 1}=1$, drawn for the case 
$N=40$ and $m=2$.}}}
\label{test}
\end{center}
\end{figure}
From this figure it is evident that the histogram matches very well with  
the Gaussian distribution $G(E)$ in Eq.~(\ref{b2a}), where
 the values of $\mu$ and $\si^2$ are obtained by using  
Eqs.~(\ref{d4a}) and (\ref{d4b}) respectively.  
Inserting the numerical values of $D^{(m)}(E_i)$ in Eq.~(\ref{d1}), 
subsequently we calculate the cumulative level
 density distribution $F(E)$  for the above mentioned case.
In Fig.$2$ we plot $F(E)$ at equidistant points 
 and also the error function $C(E)$ given in Eq.~(\ref{d2}).
\begin{figure}[htb]
\begin{center}
\resizebox{100mm}{!}{\includegraphics{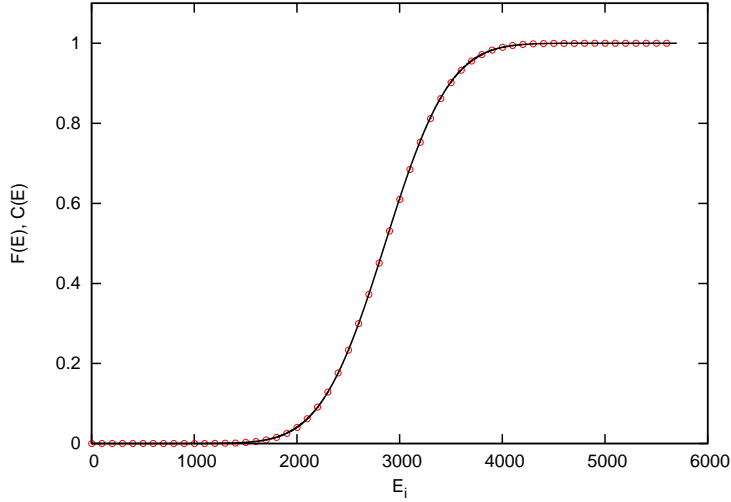}}
{\small{\caption{Continuous curve represents the error function 
$C(E)$, while the circles represent 
cumulative level density distribution  
$F(E)$ for the dispersion relation (\ref{d3}) with coefficients 
$f_{0, 1}=1$, $f_{0, 2}=-1$, $f_{1, 1}=1$, drawn for the case 
$N=40$ and $m=2$.}}}
\label{test}
\end{center}
\end{figure}
Comparison from this figure shows that $F(E)$ and $C(E)$ 
are in excellent agreement with each other. We also calculate 
the mean square error (MSE) for this case and 
find it to be as low as $4.10 \times 10^{-7}$.
In Table 1 we present the values of such MSE, which are calculated    
by taking different sets of values of 
$f_{0, 1}, f_{0, 2}, f_{1, 1}$ in the dispersion relation 
(\ref{d3}) for a wide range of $N$ (keeping the value of $m$ unchanged). 
\begin{table}[htb]
\footnotesize
\hspace{.9cm}
\renewcommand{\arraystretch}{1.9}
\begin{tabular}{|c|c|c|c|c|c|c|c|}
 \hline 
\multicolumn{3}{|c|}{\footnotesize{Sets of Parameters}}&$N=20$&$N=40$&$N=50$&$N=60$&$N=70$\\
\cline{1-3}
$f_{0,1}$&$f_{0,2}$&$f_{1,1}$&&&&&\\
\hline
$1$&$1$&$1$&$8.17 \times 10^{-6}$&$1.18 \times 10^{-6}$&$6.56 \times 10^{-7}$&$4.08 \times
10^{-7}$&
 $2.74 \times 10^{-7}$\\
 \hline
$1$&$-1$&$1$&$3.70 \times 10^{-6}$&$4.10 \times 10^{-7}$&$2.29 \times 10^{-7}$&$1.44 \times
10^{-7}$&
 $9.81 \times 10^{-8}$\\
 \hline
$1$&$2$&$1$&$1.01 \times 10^{-5}$&$1.43 \times 10^{-6}$&$7.95 \times 10^{-7}$&$4.93 \times
10^{-7}$&
 $4.93 \times 10^{-7}$\\
 \hline
$1$&$-1$&$2$&$2.46 \times 10^{-6}$&$4.09 \times 10^{-7}$&$2.33 \times 10^{-7}$&$1.48 \times
10^{-7}$&
 $1.00 \times 10^{-7}$\\
 \hline
$1$&$-1$&$3$&$3.07 \times 10^{-6}$&$5.10 \times 10^{-7}$&$2.90 \times 10^{-7}$&$1.83 \times
10^{-7}$&
 $1.24 \times 10^{-7}$\\
 \hline
\end{tabular}
\caption{ MSE for cumulative level density  
for the dispersion relation (\ref{d3}) with $m=2$}
\end{table}
From this table we find  that, for any particular choice of the three 
parameters, the MSE decreases steadily with the increase of $N$.
This fact clearly indicates that for $m=2$
the level density for the 
dispersion relation (\ref{d3}) asymptotically follows the 
Gaussian distribution. By using the same numerical method and taking 
different values of $m$, we have also studied the level densities for the  
dispersion relation (\ref{d3})
as well as a few other polynomial type   
dispersion relations and found that such level densities tend to the 
Gaussian distribution in all cases.


\vspace*{.2cm}

\no \section{Concluding remarks}
\renewcommand{\theequation}{5.{\arabic{equation}}}
\setcounter{equation}{0}
\vspace{.3cm}

By using the transfer matrix approach,  we analytically show that 
the level density for a large class of 1D vertex models 
and $Y(sl_{m})$ Yangian invariant quantum spin chains, 
associated with  polynomial type dispersion relations 
of the form (\ref{c1}), 
asymptotically follow the Gaussian distribution.
The universal nature of the asymptotic forms given 
in Eqs.~(\ref{c29}) and (\ref{c31}),
which do not depend 
on the parameters or coefficients present in the dispersion relation, 
play a key role in our analysis.
We also present some numerical evidence in support of our analytical result.
This result clearly implies that the Gaussian character
of the level density distributions for these type of models is very robust 
in nature and probably originates from the underlying Yangian quantum 
group symmetry, rather than any specific choice of the dispersion relation.
As a by product of our study on the level density distribution, 
we derive Eqs.~(\ref{c26}) and (\ref{c32}) which give general expressions 
for the order of variance and mean of the energy function respectively, 
for any polynomial type dispersion relation. 
  
It is expected that the results obtained in this article
would play an important role in investigating the 
density of spacings between consecutive energy levels of unfolded spectra 
for the class of Yangian invariant spin chains and 1D vertex models 
associated with polynomial type dispersion relations. It should be
noted that, for calculating the density of spacings,  
the energy levels are usually taken from  the unfolded spectrum which  
has an approximately uniform level density distribution \cite{H01}. 
Such unfolded spectrum can be generated from the
raw spectrum through a transformation  
involving the level density of the raw spectrum. 
Since we have shown that level density 
of all Yangian invariant spin chains and 1D vertex models 
associated with polynomial type dispersion relations
satisfy the Gaussian distribution, one can easily 
construct the corresponding unfolded spectra and analyze the 
density of level spacings for them \cite{BFG11}. 
It may also be interesting to study various thermodynamical properties
and correlation functions for this type of Yangian invariant spin chains 
and vertex models through the transfer matrix approach.



\newpage
\vspace*{1cm}
\begin{thebibliography}{99}
\providecommand{\url}[1]{{\tt #1}}
\providecommand{\urlprefix}{URL }
\providecommand{\eprint}[2][]{\url{#2}}



\bibitem{Su04} B. Sutherland, 
Beautiful Models: 70 Years of Exactly Solved Quantum Many-Body Problems
(World Scientific, Singapore, 2004). 

\bibitem{KB93} V.E. Korepin, N.M. Bogoliubov, and A.G. Izergin, 
Quantum Inverse Scattering Method and Correlation Functions,
{\em Cambridge Monographs on Mathematical Physics},
(Cambridge University Press, Cambridge, 1993).

\bibitem{EF05} F. H. L. Essler, H. Frahm, F. G\"{o}hmann,
 A. Klümper and V. E. Korepin, The One-Dimensional Hubbard Model
(Cambridge University Press, Cambridge, 2005).

\bibitem{Ha96}
Z.N.C. Ha, {Q}uantum {M}any-body {S}ystems in one {D}imension, volume~12 of
  {\em {A}dvances in {S}tatistical {M}echanics\/} (World Scientific, Singapore,
  1996).


\bibitem{Ha88}
F.D.M. Haldane, Phys. Rev. Lett. 60 (1988) 635.

\bibitem{Sh88} B.S. Shastry, Phys. Rev. Lett. 60 (1988) 639.

\bibitem{Po93} A.P. Polychronakos, Phys. Rev. Lett. 70 (1993) 2329.

\bibitem{Fr93} H. Frahm, J. Phys. A 26 (1993) L473.

\bibitem{Po94} A.P. Polychronakos, Nucl. Phys. B 419 (1994) 553.

\bibitem{FI94} H. Frahm and V.I. Inozemstsev, J. Phys. A.: Math. Gen. 27 (1994)
L801.

\bibitem{BFGR10}
J.C. Barba, F. Finkel, A. Gonz\'{a}lez-L\'{o}pez and M.A. Rodr\'{i}guez,
Nucl. Phys. B 839 (2010) 499.   

\bibitem{Ha91} F.D.M. Haldane, Phys. Rev. Lett. 67 (1991) 937.

\bibitem{Ha93} F. D. M Haldane, in Proc. 16th Taniguchi Symp., Kashikojima,
Japan (1993), eds.  A. Okiji and N. Kawakami (Springer, 1994). 

\bibitem{BGS09} B. A. Bernevig, V. Gurarie and S. H. Simon,
 J. Phys. A: Math. Theor. 42 (2009) 245206.  

\bibitem{BH09} B. A. Bernevig and F.D.M. Haldane,
Phys. Rev. Lett. 102 (2009) 066802.

\bibitem{BDS04} N. Beisert, V. Dippel, M. Staudacher. JHEP 0407 (2004) 075.

\bibitem{SS04} D. Serban, M. Staudacher, JHEP 0406 (2004) 001. 

\bibitem{BS05} N.~Beisert, M.~Staudacher, Nucl. Phys. B 727 (2005) 1.

\bibitem{BBL09} T.~Bargheer, N.~Beisert, F.~Loebbert, J. Phys. A 42 (2009) 285205.
 	
\bibitem{Se11} D.~Serban, J. Phys. A 44 (2011) 124001.  
\bibitem{HHBP92} F. D. M. Haldane, Z.N.C. Ha, J.C. Talstra, 
 D. Benard and V. Pasquier, Phys. Rev. Lett. 69 (1992) 2021. 

\bibitem{BGHP93} D. Benard, M. Gaudin, F. D. M. Haldane, and V. Pasquier, 
J. Phys.  A 26 (1993) 5219.

\bibitem{Hi95}  K.Hikami, Nucl. Phys. B 441 (1995) 530.

\bibitem{KKN97} A. N. Kirilov, A. Kuniba, and T. Nakanishi, Commun. Math.
Phys. 185 (1997) 441.

\bibitem{HBM00} K. Hikami and B. Basu-Mallick, Nucl. Phys.  B 566 (2000) 511.

\bibitem{Hi00} K. Hikami, {\it Exclusion Statistics and Chiral Partition
Function},
in ``Physics and Combinatorics 2000'', eds. A. N. Kirilov and N. Liskova, pp
22-48.

\bibitem{BBHS07}  B. Basu-Mallick, N. Bondyopadhaya, K. Hikami and D. Sen, 
Nucl. Phys. B 782 (2007) 276.

\bibitem{BBS08}  B. Basu-Mallick, N. Bondyopadhaya and D. Sen,
Nucl. Phys. B 795 (2008) 596.

\bibitem{B99} B. Basu-Mallick, Nucl. Phys. B 540 (1999) 679.

\bibitem{BMUW99} B. Basu-Mallick, H. Ujino and M. Wadati, 
Jour. Phys. Soc. Jpn. 68 (1999) 3219. 

\bibitem {BB06} B. Basu-Mallick and N. Bondyopadhaya, 
Nucl. Phys. B 757 (2006) 280.

\bibitem{Su70} B. Sutherland, J. Math. Phys. 11 (1970) 3183. 

\bibitem{Ba71} R.J. Baxter, Phys. Rev. Lett. 26 (1971) 834.

\bibitem{EFG10} A. Enciso, F. Finkel and A. Gonz\'{a}lez-L\'{o}pez,
Phys. Rev. E 82 (2010) 051117. 

\bibitem {BBK10} B. Basu-Mallick, N. Bondyopadhaya and K. Hikami, 
SIGMA 6 (2010) 091.

\bibitem {FG05} F. Finkel and A. Gonz\'{a}lez-L\'{o}pez, 
Phys. Rev. B 72 (2005) 174411.

\bibitem {EFGR05} A. Enciso, F. Finkel, A. Gonz\'{a}lez-L\'{o}pez, and M. A. 
Rodr\'{i}guez, Nucl. Phys. B 707 (2005) 553.

\bibitem {BFG09} B. Basu-Mallick, F. Finkel, and A. Gonz\'{a}lez-L\'{o}pez,
 Nucl. Phys. B 812 (2009) 402.



\bibitem {BFGR08}
J. C. Barba, F. Finkel, A. Gonz\'{a}lez-L\'{o}pez, and M. A. Rodr\'{i}guez, 
Phys. Rev. B 77  (2008) 214422.

\bibitem {BFGR08a}
J. C. Barba, F. Finkel, A. Gonz\'{a}lez-L\'{o}pez, and M. A. Rodr\'{i}guez,
Europhys. Lett. 83 (2008) 27005.

\bibitem{BFGR09}
J. C. Barba, F. Finkel, A. Gonz\'{a}lez-L\'{o}pez, and M. A. Rodr\'{i}guez,
 Nucl. Phys. B 806 (2009) 684.




\bibitem{F71} W. Feller, An Introduction to Probability Theory and 
its Applications, 3rd ed., Vol.2 (John Wiley and Sons, New York, 1971).

\bibitem{B58} N.G. de Bruijn, Asymptotic Methods in Analysis 
 (North-Holland, Amsterdam, 1958). 

\bibitem{JJ00} H. Jeffreys and B. Jeffreys, Methods
of Mathematical Physics, 3rd ed.,  (Cambridge University Press, England,
2000).

\bibitem{K76} D. Knuth, ACM SIGACT News, Vol. 8, Issue 2 (1976) 18.

\bibitem{H01}
F. Haake, Quantum Signatures of Chaos, 2nd ed. (Springer-Verlag, Berlin, 2001). 

\bibitem {BFG11} B. Basu-Mallick, F. Finkel, and A. Gonz\'{a}lez-L\'{o}pez,
 \it{Under Preparation}.
\end{thebibliography}
\end{document}